\documentstyle[preprint,aps,epsf]{revtex}
\draft
\begin{document}
\preprint{Phys. Rev. Lett. (2000) in press.}
\title{Neutron-Proton Differential Flow as a Probe of 
Isospin-Dependence of Nuclear Equation of State}
\bigskip
\author{\bf Bao-An Li\footnote{email: Bali@navajo.astate.edu}}
\address{Department of Chemistry and Physics\\
P.O. Box 419, Arkansas State University\\
State University, AR 72467-0419, USA}
\maketitle

\begin{quote}
The neutron-proton differential flow is shown to be a very useful probe 
of the isospin-dependence of the nuclear equation of state ({\rm EOS}). 
This novel approach utilizes constructively both the isospin 
fractionation and the nuclear collective flow as well as their sensitivities 
to the isospin-dependence of the nuclear {\rm EOS}. It also avoids 
effectively uncertainties associated with other dynamical ingredients of 
heavy-ion reactions at intermediate energies.
\end{quote}

\newpage
The rapid advance in experiments with rare isotopes 
has opened up several new frontiers in nuclear 
science\cite{shr93,hansen95,tanihata95,naz96,isol97,lkb98}. 
It is now possible to study in detail structures of nuclei in 
many unexplored regions of the periodic chart and novel properties 
of highly isospin asymmetric nuclear matter. Prospects for 
new physics in nuclear reactions induced by especially exotic neutron-rich nuclei 
have generated much interest in the nuclear science community. In 
particular, fast fragmentation beams at the planned Rare Isotope 
Accelerator ({\rm RIA}) will provide a unique opportunity to explore 
features of nuclear matter at extreme isospin asymmetries towards 
the pure neutron matter\cite{msu00}. The isospin-dependence of the nuclear 
equation of state ({\rm EOS}) is among the most important but very poorly 
known properties of neutron-rich matter\cite{liudo}. It is relevant 
to Type II supernova explosions, to neutron-star mergers, 
and to the stability of neutron stars. It also determines the  
proton fraction and electron chemical potential of neutron stars at 
$\beta$ equilibrium. These quantities consequently determines the cooling 
rate and neutrino emission flux of protoneutron stars and the possibility
of kaon condensation in dense stellar matter\cite{lat91,sum94,toki95,lee96}. 
At present, nuclear theories predict vastly different isospin-dependence of
the nuclear {\rm EOS} depending on both the many-body techniques and the 
bare two-body and/or three-body interactions employed. In this Letter, 
we show that the neutron-proton differential flow is a very useful probe of 
the isospin-dependence of the nuclear {\rm EOS}. This approach utilizes 
constructively both the isospin fractionation and the nuclear collective as well 
as their sensitivities to the isospin-dependence of the nuclear 
{\rm EOS}. It also avoids effectively uncertainties associated with other 
dynamical ingredients of heavy-ion reactions at intermediate
energies. The experimental measurement of neutron-proton differential flow provides
thus a novel means for determining the isospin-dependence of the 
nuclear {\rm EOS}.
 
Various theoretical studies (e.g., \cite{bom91,hub93}) have shown 
that the energy per nucleon $e(\rho,\delta)$ in nuclear matter 
of density $\rho$ and isospin asymmetry parameter
$\delta\equiv (\rho_n-\rho_p)/(\rho_n+\rho_p)$
can be approximated very well by a parabolic function
\begin{equation}\label{aeos}
e(\rho,\delta)= e(\rho,0)+S(\rho)\delta^2,
\end{equation}
where $e(\rho,0)$ is the ${\rm EOS}$ of symmetric nuclear matter. 
In the above $S(\rho)$ is the symmetry energy
\begin{equation}\label{srho2}
S(\rho)=(2^{2/3}-1)\frac{3}{5}E_{F}^{0}u^{2/3}+S_{\rm pot}(\rho),
\end{equation}
where $E_F^0$ is the Fermi energy at normal nuclear matter density $\rho_0$ 
and $S_{\rm pot}(\rho)$ is the potential contribution to the symmetry
energy. Vastly different density-dependences of $S_{\rm pot}(\rho)$ given 
by various many-body theories have led to rather divergent predictions on 
the isospin-dependence of the nuclear {\rm EOS}. Here we adopt the 
parameterization of $S_{\rm pot}(\rho)$ from the study of neutron 
stars (e.g., \cite{lat91,prak88,thor94})
\begin{equation}\label{spot}
S_{\rm pot}(\rho)=\left(S(\rho_0)-(2^{2/3}-1)\frac{3}{5}E_{F}^{0}\right)F(u),
\end{equation} 
where $u\equiv\rho/\rho_0$ is the reduced density and $S(\rho_0)$ is 
the symmetry energy at $\rho_0$. The latter is known to be in the 
range of about 27-36 {\rm MeV}\cite{farine,mass,pear}. We consider two typical 
parameterizations of $S_{\rm pot}(\rho)$ with $F(u)=2u^2/(1+u)$ and 
$F(u)=u^{1/2}$, respectively. The isospin-dependence of the nuclear {\rm EOS} 
can be characterized by the curvature of the symmetry energy at $\rho_0$ 
\begin{equation}
K_{\rm sym}\equiv 9\rho_0^2\frac{\partial^2 S(\rho)}{\partial^2 
\rho}|_{\rho=\rho_0}.
\end{equation}
For the two parameterizations of $S_{\rm pot}(\rho)$ considered here 
the $K_{sym}$ parameter is $61$ {\rm MeV} and $-69$ {\rm MeV}, respectively. 
These values are well within the wide range of $K_{sym}$ from 
about $-400$ {\rm MeV} to $+466$ {\rm MeV} predicted by many-body 
theories (e.g. \cite{lkb98,bom91}). 
Available data of giant monopole resonances does not give a 
stringent constraint on the $K_{\rm sym}$ parameter either\cite{shl93}. 
The corresponding single particle symmetry potential $V^{n/p}_{\rm asy}$ is
\begin{equation}\label{vasy1}
V_{\rm asy}^{n/p}=\pm 4e_a \frac{u^2}{1+u}\delta
+2e_a \frac{u^2}{(1+u)^2}\delta^2,
\end{equation}
and 
\begin{equation}\label{vasy3}
V_{\rm asy}^{n/p}=\pm 2e_a u^{1/2}\delta-\frac{1}{2}e_au^{1/2}\delta^2,
\end{equation}
for $K_{sym}=+61$ {\rm MeV} and $-69$ {\rm MeV}, respectively. 
In the above $e_a$ is a constant of
\begin{equation}
e_a=S(\rho_0)-(2^{2/3}-1)\,{\textstyle\frac{3}{5}}E_F^0\approx 19~ {\rm MeV}.
\end{equation}
The $+$ and $(-)$ sign is for neutrons and protons, respectively.
For small isospin asymmetries and densities near $\rho_0$ the above 
symmetry potentials reduce to the well-known Lane potential which
varies linearly with $\delta$\cite{lane}. 

Our study is based on the isospin-dependent Boltzmann-Uehling-Uhlenbeck (IBUU) 
transport model (e.g., \cite{ibuu,pak97}. In this model protons and 
neutrons are initialized according to their density distributions predicted
by the relativistic mean-field (RMF) theory\cite{ren}. The isospin-dependent 
reaction dynamics is included through isospin-dependent nucleon-nucleon scatterings 
and Pauli blockings, the symmetry potential $V_{\rm asy}^{n/p}$ and the 
Coulomb potential $V_c^p$ for protons. A Skyrme-type parameterization is 
used for the isoscalar potential. For a recent review of the IBUU model, we
refer the reader to ref.\cite{lkb98}. 

To establish the essence and to reveal the advantage of analyzing 
the neutron-proton differential flow, we first study individually the 
isospin fractionation and the isospin-dependence of nucleon collective flow. 
These phenomena in nuclear reactions induced by neutron-rich nuclei 
are very interesting in their own rights. Here we concentrate on studying 
their sensitivities to the isospin-dependence of the nuclear {\rm EOS} only. 
The isospin fractionation is an unequal partitioning of 
the neutron to proton ratio $N/Z$ of unstable asymmetric nuclear matter 
between low and high density regions. It is energetically favorable for the
unstable asymmetric nuclear matter to separate into a neutron-rich 
low density phase and a neutron-poor high density 
one\cite{muller,liko,baran,pawel,cat}. 
This phenomenon has recently been confirmed in intermediate energy 
heavy-ion experiments\cite{xu00,sjy}. 
We quantify the degree of isospin fractionation by using 
the ratio of $(N/Z)_{free}$ to $(N/Z)_{bound}$. Here $(N/Z)_{free}$ 
and $(N/Z)_{bound}$ are the saturated neutron to proton ratios of nucleons 
with local densities less (free) and higher (bound) than 
$\rho_c\equiv 1/8\rho_0$, respectively. 
The density cut-off $\rho_c$ has often been used in transport models for 
identifying approximately free nucleons quickly. The specific value of 
$\rho_c$ used here does not affect our conclusions in this work. Shown in Fig.\ 1 is 
the degree of isospin fractionation as a function of the neutron to proton 
ratio $(N/Z)_{sys}$ of the reaction system (left), impact parameter (middle) 
and beam energy (right), respectively, for reactions between several {\rm Sn} 
isotopes. It is seen that the degree of isospin fractionation increases with 
both the $(N/Z)_{sys}$ and the impact parameter, but decreases with the beam energy. 
It is very interesting to see that the degree of isospin fractionation 
is rather sensitive to the $K_{sym}$ parameter. This sensitivity is very strong 
noticing that the change of $K_{sym}$ made here is only about $1/7$ of 
its variation predicted by the many-body theories. The origin of isospin 
fractionation and its dependence on the $K_{sym}$ parameter can be easily 
understood from the density dependence of the symmetry energy. 
Since the repulsive symmetry potential 
for neutrons increases with density more neutrons will be repelled 
from high density regions to low density regions. While the opposite is true 
for protons because of their attractive symmetry potentials. Furthermore, 
the magnitude of symmetry potentials is higher for $K_{sym}=-69$ {\rm MeV} 
than for $K_{sym}=+61$ {\rm MeV} for densities less than about 
$\rho_0$. One thus expects to see a higher degree of isospin fractionation with
$K_{sym}=-69$ {\rm MeV} as shown here.
  
The analysis of collective flow has been very useful in
studying various properties of nuclear matter
(e.g.,\cite{dani85,gre86,ber88,gary,pan,res97,liko95}). How useful is it 
for studying the isospin-dependence of the nuclear {\rm EOS}? To answer 
this question we study the average in-plane transverse momentum $<p_x/N>$ 
of free nucleons as a function of their reduced rapidity $y/y_{cms}$.
Shown in Fig.\ 2 is such an analysis for the reaction of $^{124}Sn+^{124}Sn$ at 
a beam energy of 50 {\rm MeV/nucleon} and an impact parameter of 5 fm. 
To make the analysis more sensitive to the symmetry potential, the incident 
energy is chosen to be near the balance energy of the reaction system 
according to the systematics of balance energies\cite{exp9}. 
Our calculations are performed with the $K_{sym}$ parameter of 
$+61$ {\rm MeV} (upper window) and 
$-69$ {\rm MeV} (lower window), respectively. The difference between 
neutron- and proton-flow is found to depend strongly on the $K_{sym}$ 
parameter. Thus, interesting information on the isospin-dependence of the 
nuclear {\rm EOS} can also be obtained from analysing the 
collective flow of nucleons. In the case of $K_{sym}=+61$ {\rm MeV} the 
repulsive Coulomb potential dominates over the negative symmetry potential 
for generating the proton-flow. The net sum of Coulomb and symmetry potentials 
for protons is actually higher than the positive symmetry potential for 
neutrons. The proton-flow is thus higher than the neutron-flow. 
In particular, near the mid-rapidity, neutrons are still flowing to the 
negative direction because of the attractive mean filed acting on them 
while protons have already started flowing in the positive sense. 
By changing the $K_{sym}$ parameter from $+61$ {\rm MeV} to $-69$ {\rm MeV} 
the magnitude of symmetry potentials is increased but the Coulomb potential 
remains the same. Thus the difference between neutron- and proton-flow gets much 
reduced as one expects and shown in the lower window of Fig.\ 2.

It is well known that collective flow is also sensitive to both the 
curvature $K_0$ of the isoscalar part of the nuclear {\rm EOS} and 
the in-medium nucleon-nucleon cross 
sections (e.g., \cite{exp9,tsang,vd,li93,sca99,bin}).
Both quantities are still associated with some uncertainties despite 
the remarkable progress made in the last two decades.
Also the symmetry potential is normally rather small compared to the isoscalar one. 
Therefore, to study the isospin-dependence of the nuclear {\rm EOS} 
one has to select delicate observables that minimizes influences 
of the isoscalar potential but maximizes effects of the symmetry potential. 
It would be ideal if these observables can also avoid effects of other
dynamical ingredients of the reaction dynamics. We introduce here such an 
observable, i.e., the 
{\it neutron-proton differential flow}
\begin{equation}
F_{n-p}(y)=\frac{1}{N(y)}\sum_{i=1}^{N(y)}p_{x_i}\tau_i,
\end{equation}
where $N(y)$ is the total number of free nucleons at rapidity $y$ and $\tau_i=1$ 
and $-1$ for neutrons and protons, respectively. The neutron to proton ratio in
$N(y)$ is determined by the degree of isospin fractionation. In essence, this 
isospin-averaged nucleon flow combines constructively contributions of the 
symmetry potential to the collective flow of both neutrons and protons.
Simultaneously, it reduces significantly influences of the isoscalar 
potential and the in-medium nucleon-nucleon cross sections as they both act 
identically on neutrons and protons. Therefore, it uses efficiently the isospin 
effects shown in both the isospin fractionation and the nucleon collective flow. 
Shown in Fig.\ 3 is the neutron-proton differential flow as a function of 
the reduced rapidity for collisions between three {\rm Sn} isotopes at a beam 
energy of 50 {\rm MeV/nucleon}. A clear signal of the isospin-dependence of 
the nuclear {\rm EOS} is seen at both the target and projectile rapidities. 
The signal becomes stronger with the increasing isospin asymmetry of the reaction 
system as one expects. Moreover, we found that a 40\% variation of the 
$K_0$ parameter or a change by a factor of $2$ of the in-medium nucleon-nucleon 
cross sections results in a less than 7\% change in the values of $F_{n-p}(y)$. 
The neutron-proton differential flow is thus a very useful probe of the 
isospin dependence of the nuclear 
{\rm EOS}. We have also performed a study on the excitation function of 
the neutron-proton differential flow from $E_{beam}/A=30$ {\rm MeV} 
to 400 {\rm MeV} for the {\rm Sn+Sn} reactions at several
impact parameters. The signal exists clearly in the whole energy range and 
is the strongest close to the lower energy end. For a given reaction system 
involving neutron-rich nuclei, what is the best beam energy to learn the 
most about the isospin-dependence of the nuclear {\rm EOS}? We suggest 
that one selects the incident energy to be close to the balance energy 
of a reaction system with similar mass but along the $\beta$ stability valley.
It is well known that the transverse flow disappears at the balance energy 
where the attractive interactions balances the repulsive 
scatterings\cite{sto85,bert87,kro89,ogi89,sul90,pet92,hua96,ls99}. At this energy 
the non-isospin related background is the smallest for studying the 
isospin-dependence of the nuclear {\rm EOS} with the neutron-proton 
differential flow. It is useful to point out here that the well-established 
systematics of balance energies\cite{wes00} can serve as a guide in 
selecting the best beam energies for the purposes discussed above.

In conclusion, a novel means for determining the isospin-dependence of the
nuclear equation of state, i.e., the analysis of neutron-proton differential 
flow, is introduced for the first time. This approach uses constructively
both the isospin fractionation and the nuclear collective flow as well as their 
sensitivities to the isospin-dependence of the nuclear {\rm EOS}. 
It also avoids effectively uncertainties associated with other dynamical 
ingredients of heavy-ion reactions at intermediate energies.

We would like to thank C.M. Ko, W.G. Lynch, A.T. Sustich and B. Zhang 
for helpful discussions. This work was supported in part by the National 
Science Foundation Grant No. PHY-0088934 and Arkansas Science and 
Technology Authority Grant No. 00-B-14.

\begin{figure}[tbp]
\vspace{2 truecm}
\setlength{\epsfxsize=14 truecm}
\centerline{\epsffile{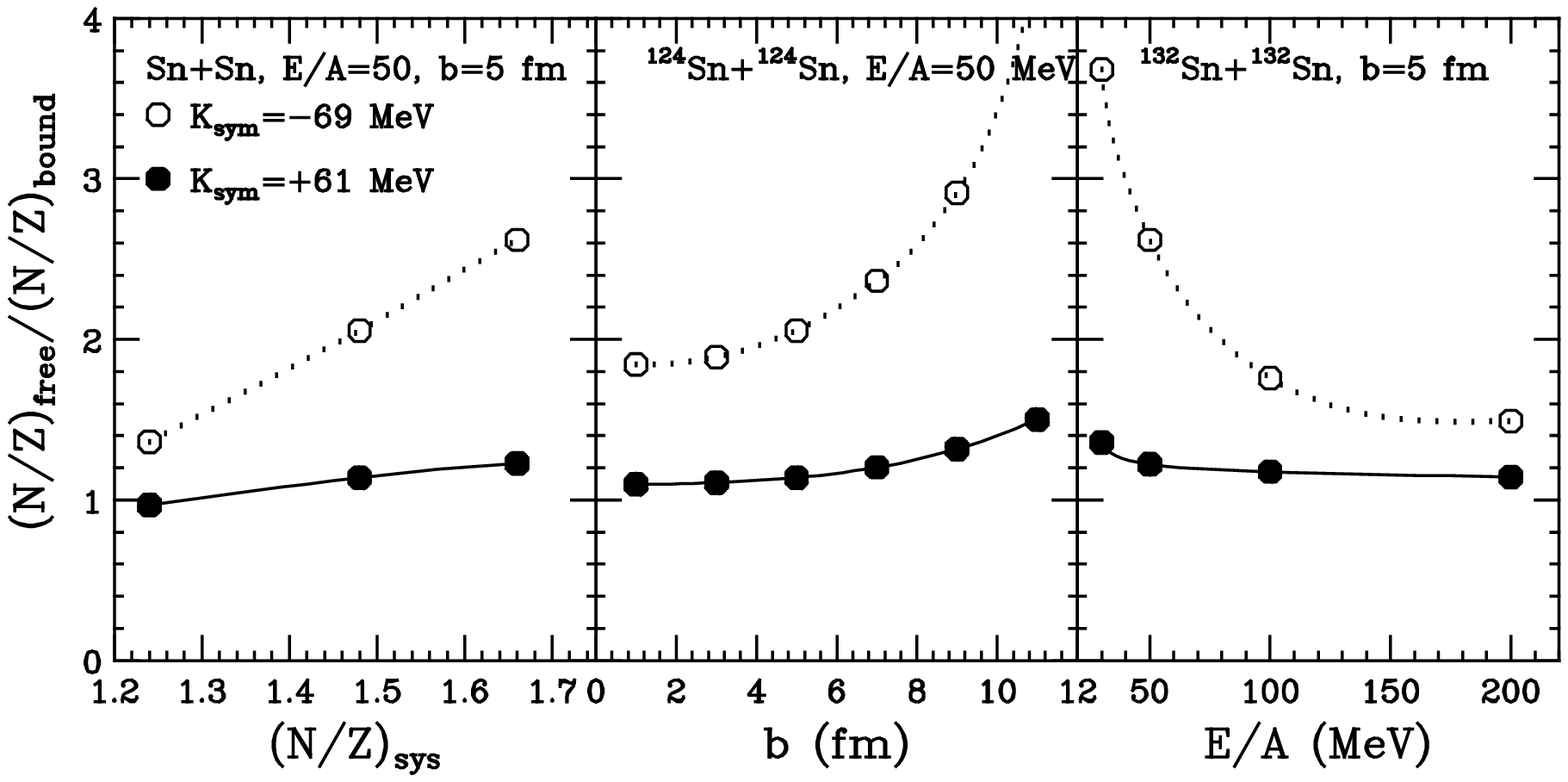}}
\vspace{-10 truecm}
\caption{Isospin fractionation as a function of (N/Z) of the rection
system (left), impact parameter (middle) and beam energy (right) in 
{\rm Sn+Sn} reactions}
\end{figure}

\begin{figure}[tbp]
\vspace{2 truecm}
\setlength{\epsfxsize=14 truecm}
\centerline{\epsffile{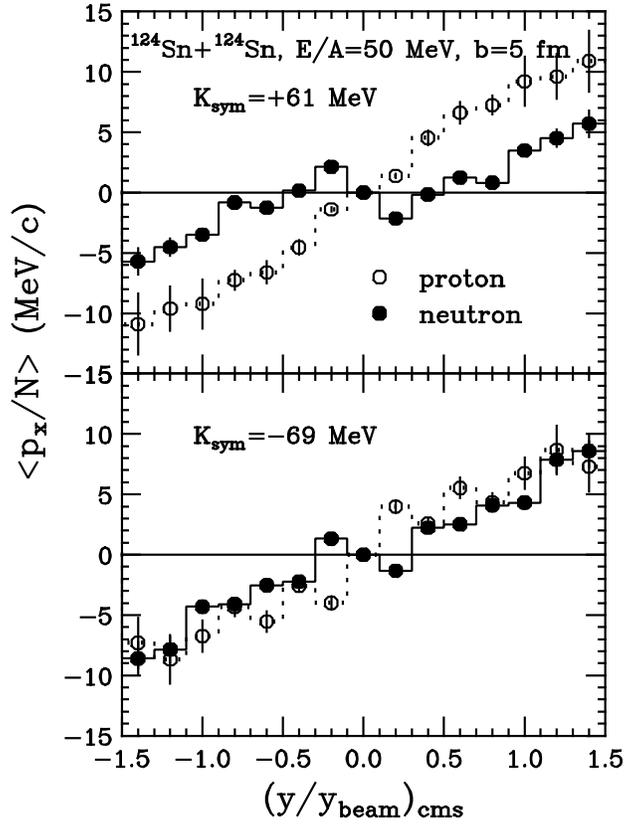}}
\vspace{-5 truecm}
\caption{The average transverse momentum per nucleon in the reaction 
plane for neutrons and protons as a function of reduced rapidity 
with the $K_{sym}$ parameter of $+61$ {\rm MeV} (upper window) and 
$-69$ {\rm MeV} (lower window), respectively.} 
\end{figure}

\begin{figure}[tbp]
\vspace{4.5 truecm}
\setlength{\epsfxsize=14 truecm}
\centerline{\epsffile{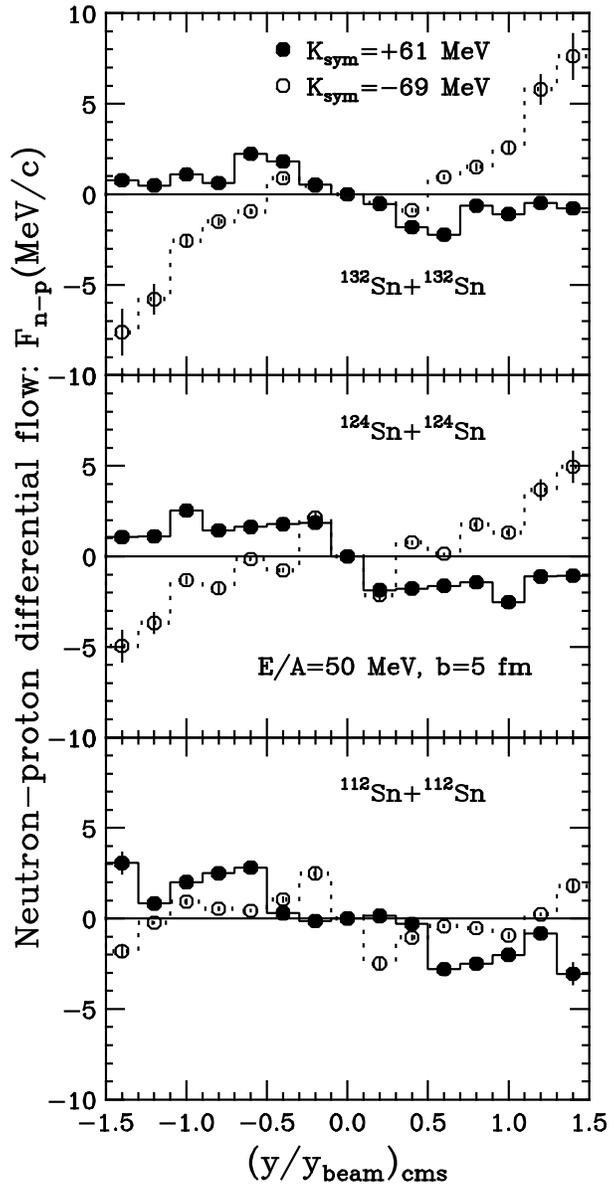}}
\vspace{-3 truecm}
\caption{Neutron-proton differential flow for the reaction of
$^{112}Sn+^{112}Sn~,^{124}Sn+^{124}Sn$ and $^{132}Sn+^{132}Sn$ at
a beam energy of 50 {\rm MeV/nucleon}, an impact parameter of 5 fm
and the $K_{sym}$ parameter of $+61$ {\rm MeV} (filled circles) and 
$-69$ {\rm MeV} (open circles), respectively.} 
\end{figure}

\end{document}